\documentclass[twocolumn,showpacs,preprintnumbers,amsmath,amssymb,prb]{revtex4}

\usepackage{graphicx}% Include figure files
\usepackage{dcolumn}% Align table columns on decimal point
\usepackage{bm}% bold math
\usepackage{color}
\usepackage{soul}
\usepackage{notes2bib}

\begin{document}

\bibliographystyle{naturemag}

%\preprint{}

\title{Optical thermometry based on level anticrossing in silicon carbide}

\author{A.~N.~Anisimov$^{1}$}
\author{D.~Simin$^{2}$}
\author{V.~A.~Soltamov$^{1}$}
\author{S.~P.~Lebedev$^{1,3}$}
\author{P.~G.~Baranov$^{1}$}
\author{G.~V.~Astakhov$^{2}$}
\email[Correspondence to ]{astakhov@physik.uni-wuerzburg.de}
\author{V.~Dyakonov$^{2,4}$}
\email[Correspondence to ]{dyakonov@physik.uni-wuerzburg.de}

\affiliation{$^1$Ioffe Physical-Technical Institute, 194021 St.~Petersburg, Russia \\
$^2$Experimental Physics VI, Julius-Maximilian University of W\"{u}rzburg, 97074 W\"{u}rzburg, Germany\\ 
$^3$St.~Petersburg National Research University of Information Technologies, Mechanics and Optics, 197101, St.~Petersburg, Russia \\ 
$^4$Bavarian Center for Applied Energy Research (ZAE Bayern), 97074 W\"{u}rzburg, Germany}

\begin{abstract}
We report a giant thermal shift of  $2.1 \, \mathrm{MHz / K}$ related to the excited-state zero-field splitting in the silicon vacancy centers in 4H silicon carbide. It is obtained from the indirect observation of the optically detected magnetic resonance in the excited state using the ground state as an ancilla. Alternatively, relative variations of the zero-field splitting for small temperature differences can be detected without application of radiofrequency fields, by simply monitoring the photoluminescence intensity in the vicinity of the level anticrossing. This effect results in an all-optical thermometry technique with temperature sensitivity of  $100 \, \mathrm{mK / Hz^{1/2}}$ for a detection volume of approximately $10^{-6} \, \mathrm{m m^{3}}$. In contrast, the zero-field splitting in the ground state does not reveal detectable temperature shift.  Using these properties, an integrated magnetic field  and temperature sensor can be implemented on the same center. 
\end{abstract}

\date{\today}

\maketitle
%---------------------------------------------------------------

Temperature sensing with high spatial resolution may be helpful for mapping of biochemical processes inside living cells and monitoring of heat dissipation in electronic circuits \cite{Yang:2011hv, Yue:2012bv, Kucsko:2013gq}.  Frequently used contact-less methods exploit temperature-dependent features either in Raman spectra of microfabricated chips \cite{Kim:2006eo, Beechem:2007fc} or in photoluminescence (PL) spectra of nanoprobes such as  quantum dots \cite{Walker:2003gc}, nanocrystals  \cite{Vetrone:2010jh, Plakhotnik:2014en} and fluorescent proteins \cite{Donner:2012gx}.  Typical temperature resolution of these methods is several hundreds of mK or lower. 

Using quantum-mechanical properties of the nitrogen-vacancy (NV) in diamond, the temperature sensitivity better than $\delta T = 10 \, \mathrm{mK / Hz^{1/2}}$ is achievable \cite{Toyli:2013cn, Neumann:2013hc, Kucsko:2013gq, Wang:2015hq}. It is based on the moderate thermal shift $d \nu_0 / d T = -74 \, \mathrm{kHz / K}$ \cite{Acosta:2010fq, Toyli:2012gl} of the optically detected magnetic resonance (ODMR) frequency in the NV center ($\nu_0 = 2.87 \, \mathrm{GHz}$ at $T = 300 \, \mathrm{K}$) and the use of the advanced readout protocols, particularly temperature-scanned ODMR \cite{Babunts:2012bu} or thermal spin echo \cite{Toyli:2013cn, Neumann:2013hc}.  However, this method is not universally usable, because the application of high-power radiofrequency (RF) fields in the pulsed ODMR technique may alter  the temperature at the probe during the measurement. Therefore, the realization of highly-sensitive and RF-free optical thermometry is of broad interest.   

Our approach is based on the silicon vacancy ($\mathrm{V_{Si}}$) centers in silicon carbide (SiC), demonstrating appealing properties for quantum sensing applications \cite{Baranov:2011ib, Riedel:2012jq, Kraus:2013vf}. Particularly, the $\mathrm{V_{Si}}$ excited state  \cite{Carter:2015vc, Simin:2016cp} shows a giant thermal shift, exceeding  $1 \, \mathrm{MHz / K}$ \cite{Kraus:2013vf}. Furthermore, these centers reveal an exceedingly long spin memory \cite{Simin:2016wv} and possess favorable absorption and PL in the near infrared spectral range \cite{Hain:2014tl}, characterized by a deep tissue penetration. The concentration of the $\mathrm{V_{Si}}$ centers can be precisely controlled over many orders of magnitude down to single defect level  \cite{Widmann:2014ve, Fuchs:2015ii} and  they can be incorporated into SiC nanocrystals as well  \cite{Muzha:2014th}.    

\begin{figure}[t]
\includegraphics[width=.48\textwidth]{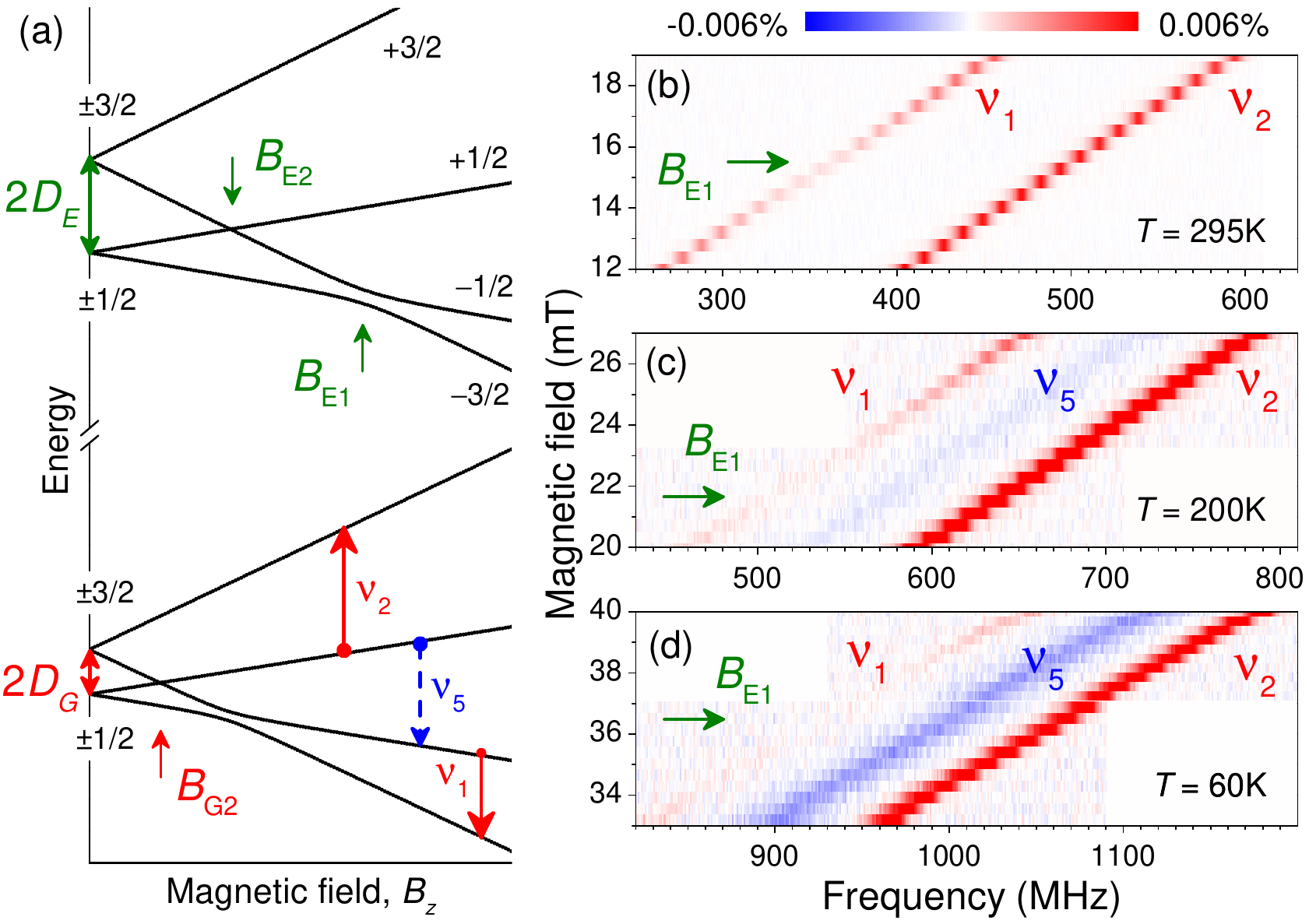}
\caption{Indirect detection of the ES spin resonance in the $\mathrm{V_{Si}}$ center of 4H-SiC.  (a) The GS and ES spin sublevels in the external magnetic field. The arrows labeled as $\nu_1$, $\nu_2$ and $\nu_5$  indicate the RF driven transitions in the GS, detected in the experiment. (b)-(d) Magnetic field dependence of the $\mathrm{V_{Si}}$ ODMR spectra  in the vicinity of the ESLAC-1, performed at different temperatures. The arrows indicate the magnetic field $B_{\mathrm{E1}}$, at which the minimum ODMR contrast of the $\nu_1$ transition is observed. } \label{fig1}
\end{figure}

We perform proof-of-concept thermometry measurements using 4H-SiC crystals.  The 4H-SiC sample under study was grown by the physical vapour transport method. Silicon vacancies were created by irradiation of the crystal with $2 \, \mathrm{MeV}$ electrons with a fluence of $10^{18} \, \mathrm{cm^{-2}}$. The  $\mathrm{V_{Si}}$ centers possess a half-integer spin state $S= 3/2$ \cite{Kraus:2013di}, which is split without external magnetic field in two Kramers degenerate spin sublevels $m_S = \pm 3/2$ and $m_S = \pm 1/2$. Here, we address the $\mathrm{V_{Si}(V2)}$ center  \cite{Sorman:2000ij} with the zero-field splitting (ZFS) in the ground state (GS) $2 D_G = 70 \, \mathrm{MHz}$ [Fig.~\ref{fig1}(a)].  The spin states are split further when an external magnetic field $B$ is applied. The spin Hamiltonian of the  $\mathrm{V_{Si}}$ center in the magnetic field has a complex form \cite{Simin:2016cp} and five RF-induced transitions are allowed: $\nu_1$  $(-1/2  \leftrightarrow -3/2)$, $\nu_2$  $(+1/2  \leftrightarrow +3/2)$, $\nu_3$  $(+1/2  \leftrightarrow -3/2)$, $\nu_4$  $(-1/2 \leftrightarrow +3/2)$ and $\nu_5$  $(+1/2  \leftrightarrow -1/2)$. In the ODMR experiments, we pump the  $\mathrm{V_{Si}}$ centers into the  $m_S = \pm 1/2$ state with a near infrared laser ($785 \, \mathrm{nm}$ or $808 \, \mathrm{nm}$ with power in the range of several hundreds mW). To decrease the detection volume to approximately $10^{-6} \, \mathrm{m m^{3}}$, we use a near-infrared optimized objective with N.A.=0.3. The PL is recorded in the spectral range from $850$ to $1000 \, \mathrm{nm}$, allowing optical readout of the $\mathrm{V_{Si}}$ spin state: it is higher for $m_S = \pm 3/2$. A detailed ODMR dependence on the magnetic field strength and orientation is presented elsewhere \cite{Simin:2015dn, Simin:2016cp}.

Due to the relatively short excited state (ES) lifetime of $6 \, \mathrm{ns}$ in the $\mathrm{V_{Si}}$ center \cite{Hain:2014tl}, the direct ODMR signal associated with the ES is weak. However, in the ES level anticrossing (LAC) between the $m_S = - 1/2$ and $m_S = - 3/2$ states (ESLAC-1)  [magnetic field $B_{\mathrm{E1}}$ in Fig.~\ref{fig1}(a)] the optical pumping cycle changes \cite{vanOort:1991ik,Martin:2000ft,Epstein:2005fi, Rogers:2009hn}. This results in a reduction of the ODMR  contrast of the corresponding GS spin resonance \cite{Carter:2015vc, Simin:2016cp}. 

Indeed, such a behavior is observed in our experiments. Figure~\ref{fig1}(b) shows the magnetic field dependence of the ODMR spectrum in the vicinity of the ESLAC-1 at room temperature. The $\nu_1$ and $\nu_2$ lines shift linearly with magnetic field applied parallel to the symmetry axis ($B  || c$) as $\nu_{1,2} = g_{\parallel} \mu_B B / h \mp 2 D_G$ for $g_{\parallel} \mu_B B / h > 2 D_G$ with $g_{\parallel} = 2.0$ denoting the g-factor. The transition with $\Delta m_S = \pm 2$ are also allowed, but corresponding $\nu_3$ and $\nu_4$ lines appear at different frequencies and have lower ODMR contrast \cite{Simin:2016cp}. The $\nu_5$ line is not resolved because of the same population of the $m_S = - 1/2$ and $m_S = + 1/2$ states under optical pumping at room temperature \cite{Kraus:2013di}. At $ B_{\mathrm{E1}} = 15.7 \, \mathrm{mT}$, the $\nu_1$ contrast drops to nearly zero and according to Fig.~\ref{fig1}(a) the ES ZFS can be determined as  $2D_E = g_{\parallel}  \mu_B B_{\mathrm{E1}} / h $. Simultaneously, the GS ZFS is directly measured as $2D_G = (\nu_2 - \nu_1)/2$. 

We repeat the above experiment at lower temperature $T = 200 \, \mathrm{K}$  [Fig.~\ref{fig1}(c)]. One can clearly see that the magnetic field associated with the ESLAC-1 is shifted towards higher values $B_{\mathrm{E1}} = 21.8 \, \mathrm{mT}$, while the splitting between the $\nu_1$ and $\nu_2$ ODMR lines remains the same. In addition, another spin resonance with negative contrast becomes visible  $\nu_{5} = g_{\parallel} \mu_B B / h$. We ascribe the appearance of the $\nu_{5}$ line with lowering temperature with different transition rates to the $m_S = - 1/2$ and $m_S = + 1/2$ states. This may occur due to the either temperature-dependent interaction with phonons or some magnetic field misalignment, which in turn leads to the modification of the intersystem crossing as well as of the optical pumping cycle. The detailed analysis is beyond the scope of this work. 

\begin{figure}[t]
\includegraphics[width=.48\textwidth]{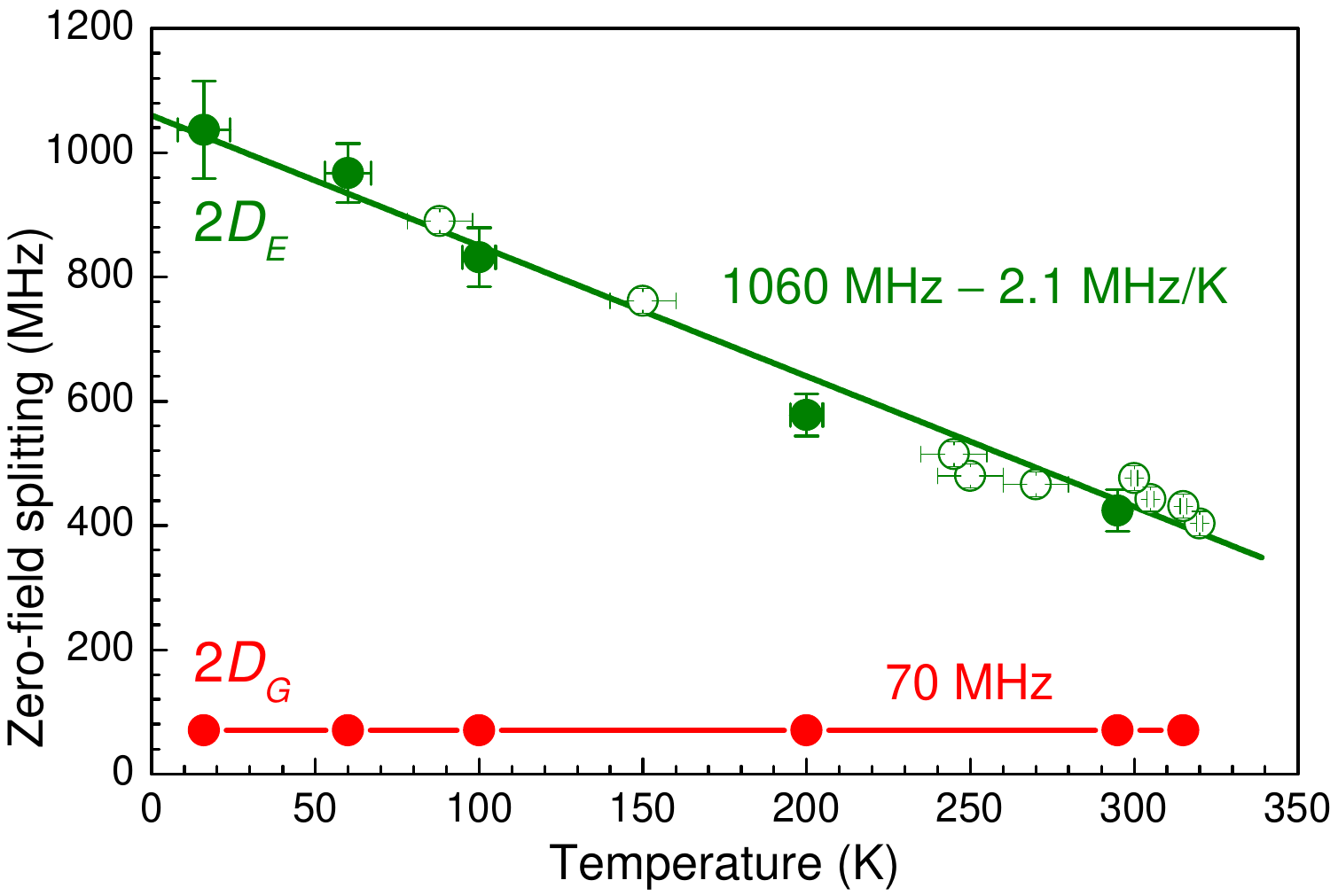}
\caption{The GS ($2D_G$) and ES ($2D_E$) ZFS in the $\mathrm{V_{Si}}$ center of 4H-SiC as a function of temperature. Solid symbols are observed from the ODMR experiments of Fig.~\ref{fig1} and open symbols from the LAC experiments of Fig.~\ref{fig3}. The line for $2D_E$ is a fit to Eq.~(\ref{Tshift}). The line for $2D_G$ is to guide the eye. } \label{fig2}
\end{figure}

The tendency continues with lowering temperature down to $T = 60 \, \mathrm{K}$  [Fig.~\ref{fig1}(d)]. Namely, we observe that the magnetic field associated with the ESLAC-1 is shifted to $B_{\mathrm{E1}} = 36.5 \, \mathrm{mT}$, indicating a further increase of $D_E$. The splitting between the $\nu_1$ and $\nu_2$ ODMR lines remains unchanged, suggesting $D_G$ is nearly temperature independent.  These findings are summarized in  Fig.~\ref{fig2}. The ES ZFS is well fitted to 
\begin{equation}
2 D_E (T)  =   2 D_E^{(0)} +  \beta T \,,  
\label{Tshift}
\end{equation}
with $2 D_E^{(0)} = 1.06 \pm 0.02 \, \mathrm{GHz}$ denoting the ZFS  in the limit $T \rightarrow 0$ and $\beta = - 2.1 \pm 0.1 \, \mathrm{MHz / K}$ being the thermal shift. The latter is by more than one order of magnitude larger than that for the NV defect in diamond \cite{Acosta:2010fq} and by a factor of two larger than previously reported for 6H-SiC \cite{Kraus:2013vf}. In following, we use this giant thermal shift for all-optical temperature sensing. 

\begin{figure}[t]
\includegraphics[width=.48\textwidth]{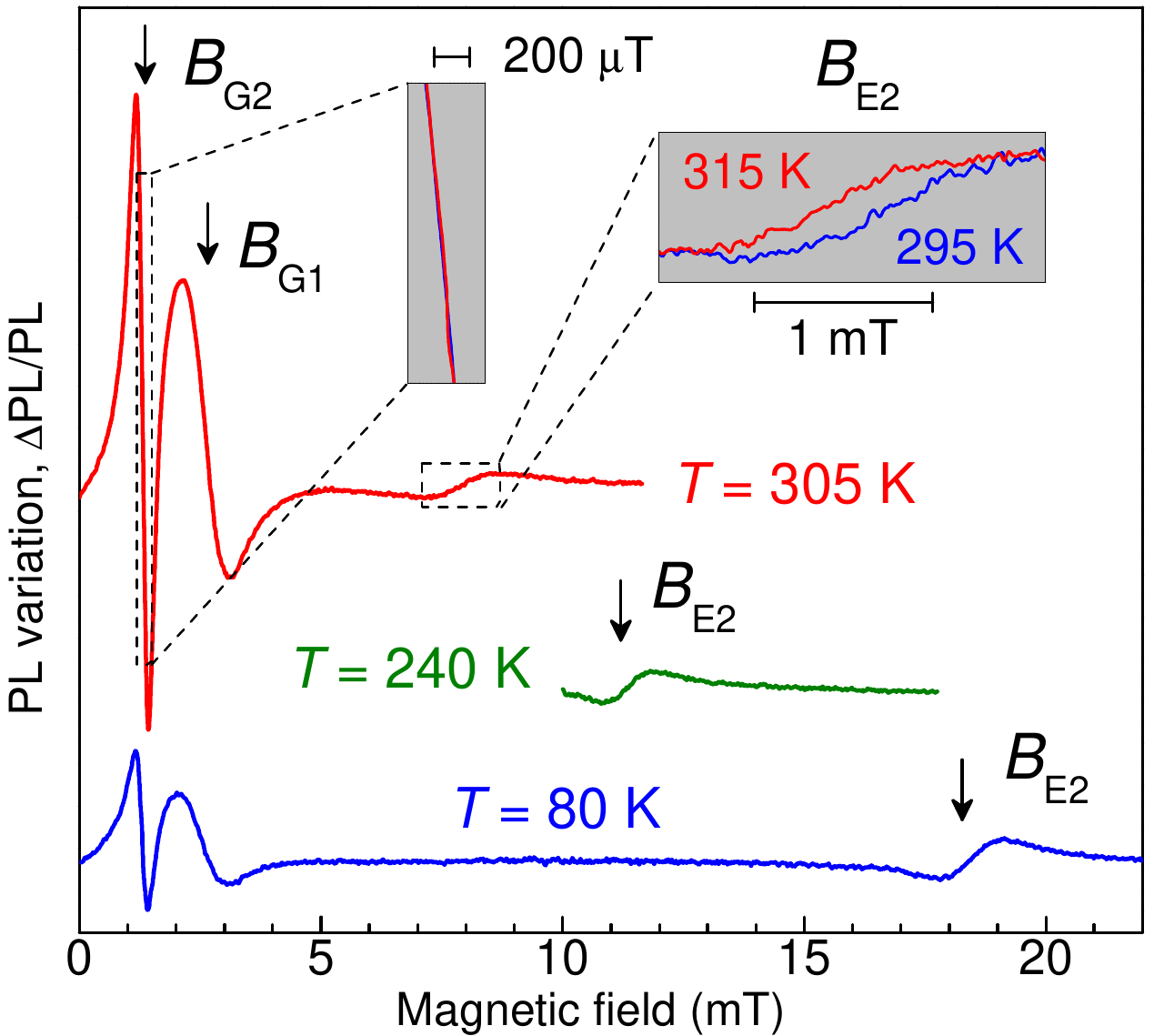}
\caption{Lock-in detection of the PL variation $\mathrm{\Delta PL / PL}$ (in-phase voltage $U_X$ normalized to the $dc$ photovoltage) as a function of the $dc$ magnetic field $B$, recorded at different temperatures. $\mathrm{\Delta PL}$ is caused by the application of an additional weak oscillating magnetic field. The arrows indicate the characteristic magnetic fields of different LACs. RF is not applied.} \label{fig3}
\end{figure}

The idea is to exploit the variation of the PL intensity in the vicinity of LAC, occurring even without RF fields. This method has been initially implemented for all-optical magnetometry in SiC \cite{Simin:2016cp}, and later extended to the NV centers in diamond \cite{Wickenbrock:2016vg}. Figure~\ref{fig3} presents lock-in detection of the PL variation $\mathrm{\Delta PL / PL}$ as a function of the $dc$ magnetic field $B_z$, recorded at different temperatures. The modulation of PL is caused by the application of an additional weak oscillating magnetic field $B$, i.e., $B_z + \tilde{b} \cos \omega t$ with $ \tilde{b} = 100 \, \mathrm{\mu T}$ and $\omega / 2 \pi = 0.33 \,\mathrm{ kHz}$. The sharp resonance at $1.25 \, \mathrm{mT}$ corresponds to the LAC between the spin sublevels $m_S = -3/2$ and $m_S = +1/2$ ($\Delta  m_S = 2$) in the GS, labeled as GSLAC-2 in Fig.~\ref{fig1}(a). A broader resonance at the double magnetic field of $2.5 \, \mathrm{mT}$ corresponds to the LAC between the spin sublevels $m_S = -3/2$ and $m_S = -1/2$ ($\Delta  m_S = 1$) and labeled, correspondingly, as GSLAC-1. The magnetic fields corresponding to the LACs in the GS ($B_{G1}$ and $B_{G2}$) are temperature independent, which is in agreement with our ODMR experiments of Fig.~\ref{fig1}. 

In addition to that, the experimental data of Fig.~\ref{fig3} reveal another resonance at the magnetic field $B_{E2}$. It corresponds to the LAC with $\Delta  m_S = 2$ in the ES (ESLAC-2), as graphically explained in Fig.~\ref{fig1}(a). Due to the strong reduction of the ES ZFS with growing temperature, this resonance shifts rapidly following Eq.~(\ref{Tshift}) as $B_{E2} = h D_E (T) / (g_{\parallel}  \mu_B)$.  We recall that the lifetime of the spin centre in the ES is about $6 \, \mathrm{ns}$ \cite{Hain:2014tl}. In order to observe ODMR signal associated with a spin state possessing such a short lifetime, one needs a RF field of about $2 \, \mathrm{mT}$. This alternating magnetic field without strong impact on the temperature of the object under measurement is difficult to achieve.

We now discuss how small variations of the magnetic field $\Delta B$ and temperature $\Delta T$ can be measured. The in-phase  lock-in voltage $U_X$ at the bias field $B_{G2}$ can be written as  (left inset of Fig.~\ref{fig3})
\begin{equation}
U_X^{G2}  =  L_{11} \Delta B +   L_{12} \Delta T \,.  
\label{U_G2}
\end{equation}
Using calibration from our earlier experiments \cite{Simin:2016cp}, we obtain $L_{11} = - 39 \,\mathrm{ \mu V / \mu T}$. Because $B_{G2}$ is temperature independent and the variation of the signal amplitude  for $| \Delta T | < 10  \, \mathrm{K}$ is negligible, $L_{12} \approx 0 \,\mathrm{ \mu V / K}$ is a good approximation. The linear dependence of Eq.~(\ref{U_G2}) holds for $| \Delta B | < 100  \, \mathrm{\mu T}$.  The same can be written for $U_X$ at the bias field $B_{E2}$ (right inset of Fig.~\ref{fig3})
\begin{equation}
U_X^{E2}  =  L_{21} \Delta B +   L_{22} \Delta T \,,  
\label{U_E2}
\end{equation}
and we find $L_{21} =  1.8 \,\mathrm{ \mu V / \mu T}$ and $L_{22} = 23 \,\mathrm{ \mu V / K}$. From the factors $L_{ij}$, it can be clearly seen that the magnetic field and temperature can be separately measured using GSLAC-2 and ESLAC-2. Particularly, the temperature sensing can be done in two steps. First, the bias field $B_{G2}$ is applied and one measures $U_X^{G2}$ to determine the actual magnetic field,  accounting for $\Delta B$ in Eq.~(\ref{U_E2}). Then, after applying $B_{E2}$ and reading out $U_X^{E2}$, the magnetic noise can be excluded from the thermometry signal using
\begin{equation}
\Delta T =   \frac{1}{L_{22}}  \left( U_X^{E2} -  \frac{L_{21}}{L_{11}} U_X^{G2} \right) .  
\label{Bration}
\end{equation}
The dynamic temperature range of such thermometry is $| \Delta T | < 10  \, \mathrm{K}$. A broad range thermometry can be realized  (with lower sensitivity) by scanning the magnetic field from $5 \, \mathrm{mT}$ to $20 \, \mathrm{mT}$ and determining $B_{E2}$, which can be then converted to temperature using $D_E = g_{\parallel}  \mu_B B_{\mathrm{E2}} / h $ in combination with Eq.~(\ref{Tshift}). 

We measure the in-phase and quadrature lock-in signals as a function of time to determine the upper limit of the noise level $\delta U$ at a given modulation frequency ($0.33 \,\mathrm{kHz}$). Then using the calibrated values for the $L$-matrix, we recalculate the noise level into the temperature sensitivity $\delta T = \delta U / L_{22}$. It is estimated to be $\delta T \approx 100 \, \mathrm{mK / Hz^{1/2}}$ within a detection volume of approximately $10^{-6} \, \mathrm{m m^{3}}$. By improving the excitation/collection efficiency and increasing the PL intensity (the $\mathrm{V_{Si}}$ concentration), the temperature sensitivity better than $\delta T \approx 1 \, \mathrm{mK / Hz^{1/2}}$ is feasible with a sensor volume of $1 \, \mathrm{m m^{3}}$. The suggested all-optical thermometry can be realized using various color centers in different SiC polytypes \cite{Falk:2013jq, Soltamov:2015ez}. Furthermore, because color centers in SiC can be electrically driven \cite{Fuchs:2013dz} even on single defect level \cite{Lohrmann:2015hd}, an intriguing perspective is the implementation of  a LAC-based thermometry with electrical readout  using photoionization of the ES \cite{Bourgeois:2015ke}.

%***********************************
%\bibliography{SiC-Thermometry}

%***********************************

\section*{Acknowledgments}
%\begin{acknowledgments}
A.N.A, V.A.S and P.G.B acknowledge support by the RSF Nr.~14-12-00859; the RFBR Nr.~14-02-91344;  16-02-00877. D.S and G.V.A acknowledge support by the BMBF under the ERA.Net RUS Plus project "DIABASE". V.D. acknowledges support by the DFG (DY 18/13-1).
%\end{acknowledgments}

\end{document}